\documentclass[%
onecolumn,
superscriptaddress,
 amsmath,amssymb,
 aps%prl
]{revtex4-2}
\usepackage[colorlinks=true, citecolor=blue, linkcolor=blue, urlcolor=blue]{hyperref}

\usepackage{graphicx}% Include figure files
\usepackage{bm}% bold math

\usepackage{wrapfig}
\usepackage{amssymb,amsmath}
\usepackage{enumitem}
\usepackage{xcolor}
\usepackage{subcaption} 
\usepackage{color,fdsymbol}

\usepackage{bibunits}

\makeatletter
\newcommand*{\newbibstartnumber}[1]{%
  \apptocmd{\thebibliography}{%
    \global\c@NAT@ctr #1\relax
    \addtocounter{NAT@ctr}{-1}%
  }{}{}%
}
\makeatother

\begin{document}

%\preprint{APS/123-QED}
%\linenumbers

\title{The mechanics of the squash nick shot}% Force line breaks with \\

\author{Mithun Ravisankar}
%\thanks{These authors contributed equally to the work}
\affiliation{School of Engineering, Brown University, Providence, RI 02912, USA}

\author{Madeline Federle}
%\thanks{These authors contributed equally to the work}
\affiliation{School of Engineering, Brown University, Providence, RI 02912, USA}
\author{Mina Bahadori}
%\thanks{These authors contributed equally to the work}
\affiliation{School of Engineering, Brown University, Providence, RI 02912, USA}

\author{Asimanshu Das}
%\thanks{These authors contributed equally to the work}
\affiliation{School of Engineering, Brown University, Providence, RI 02912, USA}

\author{Haneesh Kesari}
\affiliation{School of Engineering, Brown University, Providence, RI 02912, USA}

\author{Roberto Zenit}
\affiliation{School of Engineering, Brown University, Providence, RI 02912, USA}
\date{\today}

\begin{abstract}
Squash is a widely popular racket sport, practiced by millions of people worldwide, played inside a walled court. When played well, players can last for several minutes before the ball bounces twice on the floor. There is, however, an unanswerable shot. When the ball hits the nick between a vertical wall and the floor, under certain conditions,  it rolls without any vertical bounce. We study this process experimentally. We determined that the ball must hit the vertical wall first at a narrow range of heights above the floor, but most importantly, it must touch the floor before finishing its rolling time on the vertical wall. When the rolling time is shorter than the contact time, the vertical momentum is canceled due to a mechanical frustration condition.  This behavior is explained considering a contact model, which agrees with the experimental observations.  We argue that this concept could be relevant to the design of rolling shock dampers with many possible practical applications.
\end{abstract}

\maketitle

\section{Introduction}
Squash is a racket sport played in a four-walled court between two players. Players take turns hitting the ball in allowable areas on the walls such that the opponent is unable to return the ball back before bouncing twice on the floor. According to the Professional Squash Association \cite{PSA}, squash is played by 3 million people in over 150 countries. The sport was recently approved to be included in the 2028 Summer Olympics games in Los Angeles.

The pace, speed and court size of racket sports is determined by the mechanics of the ball \cite{cross2014,cohen2016}. What distinguishes  squash from other racket sports is the ball and its interaction with the walls. The ball is a soft-deformable hollow rubber shell that `squashes' significantly when it impacts the wall (hence the name); it is characterized by a low coefficient of restitution~\cite{cross2000,lewis2011,haron2012,cross2022}. Professional players become incredibly skilled at retrieving the ball from all parts of the court making rallies long and physically demanding. One shot in squash highly sought-after because it is unanswerable. When the ball hits the nick (the corner between a vertical wall and the floor), in some conditions, it rolls on the floor instead of bouncing~\cite{PSA_nick_shots}. This shot occurs roughly 8.7 \% in professional games (see Supplemental Information, SI): hence, it plays a significant role in the outcome of a match. From our observations, we determined that nick shots occur when the ball first hits one of the vertical walls (side or back) moving approximately perpendicular to it (see Fig. \ref{fig:1}(a)), descending at angles of incidence, $\theta_0$, ranging from 13 to 40 degrees, and at a range of different speeds. To our knowledge, no mechanical interpretation of this peculiar behavior exists. % Hern\'andez-S\'anchez and Zenit \cite{hernandez2012} studied the process, using high speed videography; they found that the ball had to hit the vertical wall first, at a certain distance away from the corner. They did not offer a mechanical interpretation of their observations.

Experiments were conducted  to identify the set of conditions needed to observe a rolling nick shot. We rationalize our measurements by placing the results on a map of the ratio of the characteristic times (contact to rolling) and the dimensionless elasticity of the ball (the Cauchy number) for collisions near a nick. We conclude by offering an explanation of the behavior sustained on a mechanical model that predicts the conditions at which the ball ceases to bounce vertically after the interaction with the nick. Understanding the mechanics of the nick shot could be relevant beyond squash and sports mechanics.  We envision using this knowledge to design rotational dampers that could be applied to sport injury prevention \cite{rowson2009,rowson2012,Fanton2019} and manufacturing control \cite{jaisee2021,saeed2023}. 

%Considering a sample of ten professional games (men and women) from the J.P. Morgan Tournament of Champions 2024, we determined that an average of 8.7\% of points are won by nick shots. 

%Using a video from PSA \cite{PSArollers}, we determined that the angles at which the nicks  occur in professional games range from 13 to 40 degrees with a mean value of $\bar\theta_o=24.8\pm 7.9$ degrees. The nicks occurred for a range of speeds.

%Nick shot description. Mostly in the front of the court where the ball motion is mostly perpendicular to one of the side walls. It can also happen on the back wall. 

\section{Experiments}

\begin{figure}[t!]
    \centering
    \includegraphics[scale=0.415]{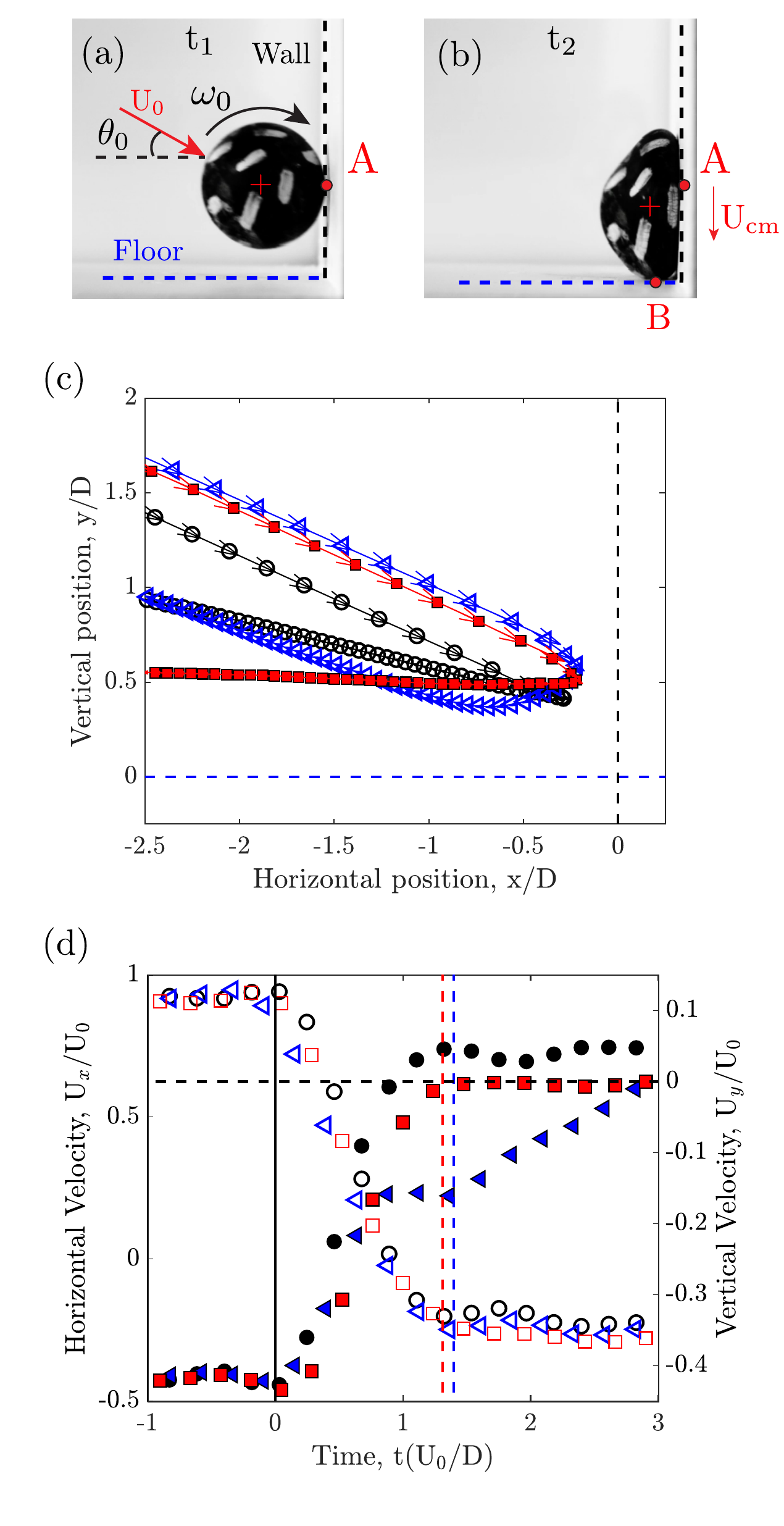}
    \caption{(a) Nomenclature, ball first contact with vertical wall, $t=t_1$. (b) Ball first contact with floor, $t=t_2$.  (c) Trajectory of the ball for three cases: Case C1, filled red squares, nick shot, Ca = $0.25$, $H^* = 0.62$, $\theta_0 = 24.43$\textdegree \ and $ \tau = 0.53$; Case C2, empty black circles, ball hits wall and floor simultaneously, Ca = $0.24$, $H^* = 0.51$, $\theta_0 = 24.34$\textdegree \ and $\tau = 0.76$; Case C3, empty blue triangles, ball hits vertical wall at higher $H$, Ca = $0.25$, $H^* = 0.81$; $\theta = 24.31$\textdegree \ and $\tau = 1.10$. (d) Normalized normalized ball velocity ($U_x/U_0$ and $U_y/U_0$) as function of normalized time, $tU_0/D$. Empty and filled symbols show horizontal and vertical velocities, respectively (left and right vertical axes, respectively). Case C1, red symbols; C2, black symbols;  C3 blue symbols; vertical solid black line corresponds to time $t_1 = 0$, vertical red dashed line shows $t_2$ for case C1; vertical dashed blue line shows $t_2$ for case C3.}
    \label{fig:1}
\end{figure}

Using the experimental setup described in the Methods Section and the SI, controlled collisions were produced. The ball was directed to hit the nick between a vertical and a horizontal wall at different locations, $H$, speeds, $U_0$ and incidence angles, $\theta_0$, see Fig. \ref{fig:1}(a). Experiments were conducted for three different balls, two different ball temperatures and two fixed incident angles. All experiments consider no initial rotational speed, $\omega_0=0$. Table \ref{Table:ExperimentalConditions} summarizes all the experimental conditions considered.   

% \begin{figure}[t]
%	\centering
%		\includegraphics[width=0.5\textwidth]{Fig_1-01.eps}}
%	\caption{Schematic of the experimental setup for reproducing the nick shot. An acrylic corner wall is used for the ball to hit. The ball is launched using a air pressurized canon (not shown here), and images are recorded using a high-speed camera. The soft ball undergoes significant deformation as shown on the right image. Circular speckle spots are painted on the ball to allow for tracking the rotation of the ball.}
%	\label{fig1:Expt_setup}
%\end{figure}

%The interaction process can be characterized as follows. The ball, with mass $m$, diameter $D$, and material elasticity $E$, approaches the corner at a speed $U_o$ and rotational speed $\omega_o$, at an incident angle $\theta_o$, as depicted in Fig. \ref{fig:setup1}. 

%The interaction starts at time $t_1$, when the ball makes contact with the vertical wall at a height $H$ above the floor, corresponding to a material point $A$ on the ball. We assume that the point $A$ does not occur upon contact, since $\theta_o$ is small and wall-ball friction is relatively large \cite{maw1976}). %As shown above, the nick shot occurs when the ball makes contact with the floor while is still in contact with the wall: $H/D>0.5$.
We rationalize the interaction by considering the following phases: (i) contact with the wall, $t=t_1$; (ii) compression and rolling, $t_1<t<t_2$;  (iii) contact with the floor $t=t_2$; and (iv) bounce or nick process, $t>t_2$. Figures \ref{fig:1}(a) and \ref{fig:1}(b) show nomenclature and snapshots of a typical collision at $t=t_1$ and at $t=t_2$, respectively. Figures \ref{fig:1}(c) and \ref{fig:1}(d) show the trajectory and the speed time evolution, respectively, of three typical collisions (cases C1, C2 and C3, see movies in the SI). In particular, the case C1 shows the nick behavior, depicted by the red symbols in the figure. The ball first touches the vertical wall at a point $A$,  at distance $H^*=H/D>0.62$, at $t=t_1$. Immediately after, the ball slows down as it exchanges momentum with the wall. As the ball  compresses, it also rolls without sliding at the contact point $A$ \cite{maw1976} (conditions discussed in the SI). Hence, the ball's center of mass continues to move downwards (see vertical speed of the ball, $U_{cm}$, for $t_1<t<t_2)$, even though the point $A$ is fixed. As the interaction evolves, the ball eventually touches the floor on point $B$ at $t=t_2$, Fig. \ref{fig:1}(b). Soon after this  happens, the vertical velocity of the ball becomes zero. The most impressive part of the nick behavior is displayed after this moment. As the ball decompresses, it horizontal velocity gradually increases until loosing contact with the wall but, most importantly, its vertical velocity remains very close to zero. In other words, the ball leaves the corner interaction with a nearly horizontal trajectory, as seen in Fig. \ref{fig:1}(c). Two other cases are shown in the figure, for $H^*=0.51$ and $H^*=0.81$. Even though the ball collides at approximately the same speed and incident angle, the nick behavior is not observed (detailed description in SI).

%As seen in Fig. \ref{fig:setup} (f), when the ball nicks the vertical velocity is very small, maintaining a nearly constant elevation, close to the floor.. In Figs. \ref{fig:setup}(e) and (f), a case for which the nick condition is not met is also shown. In this case, the speed was nearly the same but the ball hit the vertical wall too close to the nick, reaching both vertical and horizontal walls at approximately the same time. In this case, the ball also undergoes a compression phase, but the rolling process does not occur. Many experiments were conducted to identify the range of conditions when a rolling nick was observed.

%We conjecture that the rolling nick behavior appears when the rolling process of the ball is frustrated by the horizontal wall. When this occurs, the ball gets mechanically stocked at the corner.  

\begin{table}[t!]
\caption{Range of conditions and ball properties for all experiments. For all cases, $0.5<H^*<1.0$.}
\begin{tabular}{p{0.75cm} p{4.5cm} p{1.5cm} p{2cm}  p{2cm}}
\hline
Case & Ball type & E & $U_o$  &  Ca\\
&& (kPa) & (m/s) & \\
\hline
1 & Single dot (\textcolor{blue}{$\medcircle$}, \textcolor{blue}{$\medblackcircle$}) & 99 & 33 to 58  &  0.05 to 0.12 \\
2 & Double dot  (\textcolor{blue}{$\square$}, \textcolor{blue}{$\medblacksquare$}) & 105 & 18 to  58  & 0.04 to 0.13 \\
3 & Doubles  (\textcolor{blue}{$\medtriangleup$}, \textcolor{blue}{$\blacktriangle$}) & 471 & 17 to 58 & 0.16 to 2.26 \\
4 & Single dot, at 40\textdegree \ C (\textcolor{red}{$\medcircle$}, \textcolor{red}{$\medblackcircle$}) & 96 & 30 to 52   & 0.04 to 0.15 \\
5 & Double dot, at 40\textdegree \ C (\textcolor{red}{$\square$}, \textcolor{red}{$\medblacksquare$}) & 90 & 33 to 56  & 0.04 to 0.11 \\ 
6 & Doubles, at 40\textdegree \ C (\textcolor{red}{$\medtriangleup$}, \textcolor{red}{$\blacktriangle$}) & 450 & 30 to 55 & 0.21 to 0.62 \\ \hline
\end{tabular}
\label{Table:ExperimentalConditions}
\end{table}

\subsection{Condition for a nick shot}
We hypothesize that the nick will occur if the ball makes contact with the floor before it has ended its rolling motion. In other words, the contact time has to be longer than the rolling time. We can estimate the characteristic time for the compression, $t_c$, considering the collision of of an elastic ball against a rigid wall \cite{johnson1987}. 
The rolling time, $t_r$, can be calculated by considering the displacement of the ball's center of mass during contact and the rotational inertia of the ball (details of calculations appear in the SI).

If we define $\tau={t_r}/{t_c}$, the condition for a nick shot is
\begin{equation}
\tau<1.
\end{equation}
Considering explicit expressions for $t_c$ and $t_r$, we obtain
\begin{equation}
    \tau=\beta H^*\text{Ca}^{2/5} \frac{(\cos\theta_0)^{1/5}}{\sin\theta_0+2 \kappa \omega^*} 
    \label{eqn:condition}
\end{equation}
where $\beta=0.6058$ is a numerical constant, $\omega^*= \omega _0D/U_0$, $H^*=H/D$ and $\text{Ca}$ is the Cauchy number defined as
\[\text{Ca}=\frac{E}{\rho_{ball} U_0^2}\]
where $E$ is the Young's modulus and $\rho_{ball}=6m/(\pi D^3)$, is the density of the ball.  Using  Eqn. \ref{eqn:condition}, a map of conditions for the nick to occur can be obtained, considering the collision kinematic conditions ($U_0$, $\omega_0$, $\theta_0$), the location of first contact with the wall ($H^*$), and the ball elasticity parameter (Ca).

\section{Results and discussion}
\begin{figure}[h]
    \centering
    \includegraphics[scale=0.45]{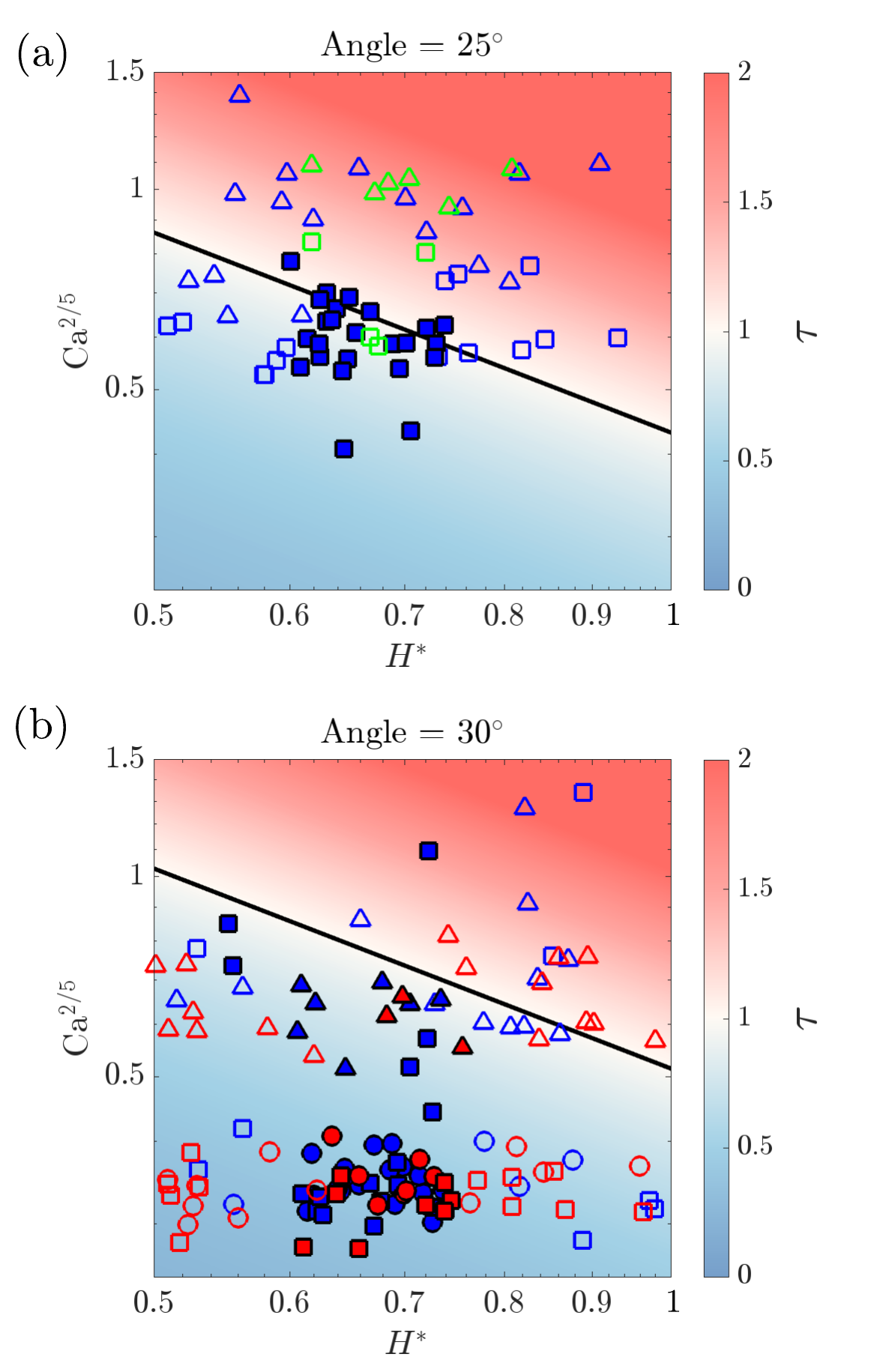}
    \caption{Map of conditions for nick occurrence: value of the time ratio $\tau$ (indicated by the color) as a function of Ca$^{2/5}$ and $H^*$ for a fixed incidence angle $\theta_0$ and no initial rotation $\omega_0=0$. (a) $\theta_0=25$\textdegree \ ; (b) $\theta_0=30$\textdegree. Symbols according to Table \ref{Table:ExperimentalConditions}. Filled  symbols (resp. empty) show the cases when the nick shot was observed (resp. was not  observed). Green symbols on (a) show experiments when the wall was lubricated with a viscous fluid.
}
    \label{fig:2}
\end{figure}

A vast experimental campaign was conducted (Table \ref{Table:ExperimentalConditions}) to test the conditions set by Eqn. \ref{eqn:condition}. In all experiments, no initial rotation was considered ($\omega^*=0$) and only two representative incident angles, $\theta_0$, where tested.  Figure \ref{fig:2} shows all the experimental results. In these maps, we plot the value of $\tau$ for a range of values $H^*$ and Ca. On the plots, the filled symbols indicate the cases nick was observed.  A shot was considered to be a nick if the rebound vertical speed was smaller than $0.1 U_0 \sin\theta_0$. Clearly, two conditions can be identified: $\tau<1$ and $0.6<H^*<0.75$.  %The agreement with the prediction of Eqn. \ref{eqn:condition} is remarkable, without any fitting parameters.
Even without any fitting parameters, Eqn. \ref{eqn:condition} predicts the conditions very well. 

Several other conditions were also tested, including ball temperature, ball type, ball speeds and wall lubrication, all of which are shown in Fig. \ref{fig:2}. A detailed account appears in the SI. The conditions for the occurrence of the nick shot hold for all these tests.

Considering a mechanical model (described in the SI), we can explain why the ball does not bounce vertically when the nick shot conditions are satisfied. The rolling process that occurs when the ball is in contact with the vertical wall cannot proceed when a second contact point, with the floor, is imposed. This second contact point induces a torque in the opposite direction of the rolling motion, canceling it, resulting in  a mechanical equilibrium condition for which both the linear and rotational momentum of the ball become null. In other words, the ball gets stuck on the corner when the nick conditions are met. Because of the energy stored during the compression phase with the vertical wall, the ball can still rebound horizontally. This is why the ball moves away from the wall after the collision with no vertical momentum.

Understanding the mechanics of the nick shot can be used to train professional players produce that shot more frequently. In addition to precision, to place the ball within the narrow range of heights $0.6<H^*<0.75$, the nick behavior to occurs for small values of $\tau$. This occurs, according to Eqn. \ref{eqn:condition}, for higher speeds (smaller Ca), higher angles of incidence or softer balls (smaller $E$). So, hit hard, from a fully extended arm and when the ball nicely warm (after a few minutes of hard hitting). Now that we understand the mechanics, we shall be invincible in the courts. 

\subsection{Materials and Methods}

Controlled collisions were produced with an air gun. It deployed squash balls (Dunlop, Professional 2-Dot, one yellow dot and elite doubles) at a range of speeds, close to those use in regular squash. The incident angle and the initial contact with the vertical wall was varied, by moving the target and its inclination. The collisions were recorded with a high speed camera. The position was determined using image processing software; the velocity was calculated using a central-difference scheme. More details can be found on the SI.

\vspace{2cm}

\textit{Acknowledgements} - We thank Mark Johnson and Alex Zaslavsky from the Rhode Island Squash Association and Stefano Crema from Squash Busters Providence for animated discussions. We are also grateful to Coach Arthur Gaskin, from Brown University, for his wisdom in all things squash.

\bibliography{reference_squash}

\newpage

\begin{center}
  {\Large Supplementary Information: The mechanics of the squash nick shot}  
\end{center}
% Force line breaks with \\

% \author{Mithun Ravisankar}
% %\thanks{These authors contributed equally to the work}
% \affiliation{School of Engineering, Brown University, Providence, RI 02912, USA}

% \author{Madeline Federle}
% %\thanks{These authors contributed equally to the work}
% \affiliation{School of Engineering, Brown University, Providence, RI 02912, USA}
% \author{Mina Bahadori}
% %\thanks{These authors contributed equally to the work}
% \affiliation{School of Engineering, Brown University, Providence, RI 02912, USA}

% \author{Asimanshu Das}
% %\thanks{These authors contributed equally to the work}
% \affiliation{School of Engineering, Brown University, Providence, RI 02912, USA}

% \author{Haneesh Kesari}
% \affiliation{School of Engineering, Brown University, Providence, RI 02912, USA}

% \author{Roberto Zenit}
% \affiliation{School of Engineering, Brown University, Providence, RI 02912, USA}
% \date{\today}

\section*{Statistics of nick shot occurrence}
To determine how often nick shots are played, one of us (RZ) watched all the matches (men and women, 32 matches) played at the Tournament of Champions of 2023, transmitted by SquashTV. This prestigious tournament, part of the professional tour, is played every January in New York City. We counted the number of times a player won a point with a nick shot. It occurred 8.7 \% times. Note that not all of these shots were rolling nicks. Those occurred only 1.2\% of all winning shots.

\section*{Experimental Setup}

The experiments were conducted in the controlled environment of our lab rather than on the squash courts. To simulate the player hitting the squash ball, a pressurized air cannon was used to launch a squash ball towards an acrylic wall. The air cannon, constructed of PVC piping with an inner diameter close to that of the diameter of the balls, a plastic chamber, a manometer, and a valve. Once pressurized to a certain value, the valve was quickly opened to release the squash ball at a certain speed $U_0$, aimed at the corner at an incident angle $\theta_0$. The pressure inside the air cannon was varied from 15 psi to 40 psi to produce a range of impact speeds, as seen in Table~1 in the main manuscript. The acrylic wall represents the nick, where the vertical wall and the floor meet. The corner, constructed from two pieces of 7.5 cm thick acrylic plate, was mounted using aluminum profiles about 1 meter away from the air cannon. The angle at which the acrylic corner was mounted was then varied from 20 to 40 degrees to produce different incident angles.  The impact location $H$ was varied by displacing the acrylic corner back and forth, while keeping the position of the cannon fixed. 

Three different squash balls  were used, a single dot (for intermediate), a double dot (for professional players), and a hard ball (used in doubles), all manufactured by Dunlop. The diameter of the balls were $D= 40$ mm, and their masses were $m$ = 24 grams.

The ball's Young's modulus was determined experimentally following the procedure described below. For the collision experiments and the compression tests, the balls were tested at room temperature, 22\textdegree C, and at 40\textdegree C, which we determined to be a typical temperature for a squash ball during a hard game. The ball was heated using a heating pad. The temperature was measured using a hand-held infrared heat gun, with an accuracy of $\pm 2$\textdegree C. 

The interaction between the squash ball and the wall was recorded using a high-speed camera (\textit{Photron FASTCAM SA5}) filming at 5000 fps, with a shutter speed of 1/20000. Examples of typical experiments can be seen in movies below. The squash ball was marked up with white stripes to help image tracking. 

The images collected were analyzed using image processing techniques on Matlab. Using the diameter of the squash ball for calibration, the location of the ball was tracked in time, as shown in Fig. 1(c) of the main manuscript. Hence, both vertical and horizontal velocities of the ball geometric center were determined during the interaction with the wall. The ball speed was determined in two directions, considering a central difference scheme. The occurrence of a nick shot was determined from these measurements.  A collision was considered a nick if the vertical velocity after the interaction was smaller that $0.1 U_0\sin\theta_0$.

\section*{Measurement of the ball's Young's Modulus}

To determine the Young's Modulus of the squash balls, compressive tests were conducted using an Instron machine. The machine was set to compress each ball at a controlled rate of 2 mm/s up to a predetermined deformation limit of 25 mm, well below the permanent deformation limit. The applied force as a function of deformation showed a non-linear behavior. We, therefore, used the two-parameter hyperelastic Gent model \cite{gent1996new,das2020nonlinear}, to determine the effective Young's modulus of the balls. The stress, $\sigma$ was calculated as the ratio of the compression force over the cross sectional area of the ball. The strain, $\epsilon$, was calculated as the ratio of the deformation and the ball diameter. The data was fitted to 
\begin{equation}
    \sigma = G_m \left ( \lambda - \frac{1}{\lambda^2}\right),
\end{equation} 
where, $\lambda=1 + \epsilon$ is the stretch ratio and $G_m$ is normalized shear modulus defined as 
\[G_m = G J_m /(J_m - I_1 + 3)\]
where  $G$ is the shear modulus, $J_m$ is the locking parameter and $I_1$ is the first invariant of the left Cauchy-Green deformation gradient tensor. The fitting of the data to this model yielded the material constants, $G$ and $J_m$. The Young's modulus was determined by
\begin{equation}
   E = 2G(1+\nu),
\end{equation} 
where $\nu=0.5$ is the Poisson ration for rubber. The values of $E$ are shown in Table 1 of the main document.

\section*{Characteristic times for wall collision}
To identified the conditions for the nick shot to occur, two characteristic times are needed: the compression time and the rolling time. Here, we detail how those times are determined.

\subsection*{Contact time}
The contact time for the impact of an elastic sphere is determined by the balance of the impact force and the elastic deformation, neglecting friction and energy dissipation \cite{johnson1987}. The impact time is give by
\begin{equation}
    t_c=\alpha\left(\frac{m^2}{D E^2U_0\cos\theta_0}\right)^{1/5}
    \label{contact_time}
\end{equation}
where $\alpha=3.29$, $m$ is the mass of the ball, $U_o$ is the ball speed at contact, $\theta_0$ is the angle at contact and $E$ is the Young's modulus. 

\subsection*{Rolling time}
Assuming that there is no slip between the  ball and the wall, a condition met for contact with friction and at small incident angles \cite{maw1976}, the ball will execute a rolling motion on the wall with in contact with it. The time that the wall will take to reach the floor, while rolling, is
\[t_r=\frac{H}{U_{cm}}\]
where $H$ is the distance above the floor where the ball first makes contact with the wall. $U_{cm}$ is the velocity of ball's center of mass at time $t_1$ (see Fig. 1(a) in the main text), which can be defined as
\begin{equation}
   U_{cm}=\frac{D(U_0\sin\theta_0-D\omega_0/2)}{D_J+D} 
\end{equation}
where $D_J=2J/(mD)$, $J$ being the ball's polar moment of inertia and $\omega_0$ is the ball's angular velocity at the moment of contact. This calculation the classic rolling theory \cite{johnson1987}. Therefore, we can rewrite $t_r$ as:
\begin{equation}
    t_r = \frac{H (2 D_J + D)}{D ( U_0 \sin \theta_0 + D_J \omega_0)}
\end{equation}
where $D_J=2J/(m D)$. A squash ball is a thick spherical shell of thickness $T$, for which the polar moment of inertial is
\begin{equation}
     J = \frac{1}{10} m D^2 \frac{1 - (1 - 2T/D)^5}{1 - (1 - 2T/D)^3}.
\end{equation}
Considering that $T/D=0.113$ for a squash ball, $J=\kappa m D^2$ where $\kappa=0.1346$.

Therefore, the rolling time simplifies to: 
\begin{equation}
    t_r=\frac{H(4\kappa + 1 )}{U_0\sin\theta_0+ 2 \kappa D \omega_0}
    \label{rolling_time}
\end{equation}

\subsection*{The time ratio, $\mathbf{\tau}$}
The ratio of rolling to contact time is defined as
\[\tau=\frac{t_r}{t_c}.\]
Therefore, from Eqns. \ref{contact_time} and \ref{rolling_time}, we can write
\[\tau= 
\frac{1}{\alpha} \left(\frac{D E^2U_0\cos\theta_0}{m^2}\right)^{1/5}
\frac{H(4\kappa + 1 )}{U_0\sin\theta_0+ 2 \kappa D \omega_0}\]

After simplification, we have
\begin{equation}
    \tau=\beta H^*\text{Ca}^{2/5} \frac{(\cos\theta_0)^{1/5}}{\sin\theta_0+2 \kappa \omega^*} 
    \label{eqn:condition}
\end{equation}
where $\beta=0.6058$ is a numerical constant, $\omega^*= \omega _0D/U_0$, $H^*=H/D$ and 
\[\text{Ca}=\frac{E}{\rho_{ball} U_0^2}\]
where $\text{Ca}$ is the Cauchy number and $\rho_{ball}=6m/(\pi D^3)$.

\section*{The mechanical model of the ball contact}

\subsection*{Stage 1: Contact with the wall, $t = t_1$}
Before making contact with the vertical wall, the ball travels at a velocity $\mathbf{U_0}$, approaching the wall at an angle $\theta_0$, as depicted in Fig. 1(a) of the main text. When the ball reaches the wall, its horizontal location is $x/D=0.5$ while its vertical location, $H^*=y/D$, depends in how close to the floor it reaches the vertical wall. The ball makes contact with the wall at a material point on its surface, denoted as $A$. After making contact, the ball experiences an impulse which changes its center of mass velocity, as well as its angular velocity. From the balance of linear and angular momentum, the change of the ball's center of mass velocity is given by
\[\Delta {U_{cm}}=D_J \Delta \omega\]
where $\Delta \omega$ is the change of angular velocity before and after the impact and $D_J=2J/(mD)$, with $J$ being the polar moment of inertia (see Eqn. 5 above).

Assuming that the velocity of a material point of the ball at location $A$ (no slip) is zero, we can conclude that
\[U_0 \sin \theta_0 + \Delta U_{cm} - \omega D/2 = 0\]
and that
\[\Delta \omega=\frac{\tfrac{1}{2}U_0 \sin \theta_0-\omega_0 D}{D_J+D}.\]
Therefore, the speed of the center of mass immediately after contact is
\begin{equation}
   U_{cm}=\frac{D(U_0\sin\theta_0-D\omega_0/2)}{D_J+D}.
\end{equation}

\subsection*{Stage 2: contact with the floor, while already in contact with the wall, $t=t_2$}
As described in the main text, when the ball has not yet concluded its rolling motion on the vertical wall but makes contact with the floor, the nick behavior is manifested. In this case the ball is observed to roll on the floor with no vertical velocity.

When the ball makes simultaneous contact with the vertical wall, at point $A$ with the vertical wall and at point $B$ with the floor, it experiences impulses from these two points that will change the ball's velocity. Let the change of the ball's center of mass at $t=t_2$ be
\[\Delta U_{cm}(t_2)=P_B/m+P_A/m\]
where $P_B$ and $P_A$ are the impulses at $B$ and $A$, respectively.

At point $A$, considering that at $t=t_2$ the material point is still fixed (non slip), we will have
\begin{equation}
    U_{cm}(t_2)=D \omega/2.
    \label{eqn:cond1}
\end{equation}
Now, we can also assume that at the material point $B$ the the velocity is null. Therefore, following a similar argument as before, we can conclude that
\begin{equation}
    U_{cm}(t_2)=-D\omega/2.
    \label{eqn:cond2}
\end{equation}
The only possibility for Eqns. \ref{eqn:cond1} and \ref{eqn:cond2} to be both satisfied is for $\omega=0$. From it, it follows that $U_{cm}=0$.

\subsection*{Stage 3: restitution}
As shown above, if the ball makes simultaneous contact with the vertical wall and floor, while rolling on the vertical wall, its velocity and rotation cancels out. However, due to the impact, the ball still has energy stored in elastic deformation. Since the energy is mostly from deformation in the horizontal direction, the ball will regain kinetic energy in this direction. Due to the contact at point $B$, the ball will roll as the elastic energy is restored into kinetic energy in the horizontal direction.

\section*{Detailed description of experiments}

\subsection*{Experiments when the nick shot is not observed}

In this section we provide a detailed description of two experiments shown in Fig. 1(c) and 1(d) (case C2 and C3) that do not exhibit the nick behavior.

Our observations indicate that the nick behavior occurs when the balls touches the floor before it has completed the rolling phase with the vertical wall. To illustrate that this is indeed the nick condition, we show two other cases in Fig.1 of the main document, with very similar conditions but reaching the corner at slightly different $H^*$ values. Case C2, shown in Figs. 1(c) and 1(d) by the black markers and the movie below, displays a collision for which $H^* = 0.51$; in this case, the ball makes contact with both vertical wall and floor at approximately the same time ($t_1\approx t_2$). The compression-decompression cycle occurs with both walls at the same time and rolling does not occur. As explained above, rolling is not possible while having two no-sling contact points.
The ball bounces off with at a slightly lower angle that the incoming one, see Fig. 1(d). Case C3, displayed by the blue symbols, also in Figs. 1(c) and 1(d) and movie below,  shows an interaction for which $H^* = 0.81$. The collision occurs at a higher location than C1 (nick case). The evolution of the ball position and velocity is very different. The evolution of position and speed, for times soon after $t=t_1$ occurs in a manner similar to Case C1: the ball compresses and rolls on the vertical wall. Since the vertical location is higher, the ball is capable of completing is rolling motion with respect to point $A$ by the time it touches the floor at time $t_2$. Hence, the ball is free to continue its vertical motion downwards, starting a compression phase with the floor. During this compression phase with the floor, the ball is not longer in contact with the wall on point A. Therefore, the motion of the center of mass of the ball has a negative velocity even for times longer that $t_2$ even as the ball is already moving away from the wall in the horizontal direction, as seen on Fig. 1(d). Subsequently, the ball bounces off from the floor with a certain angle, with both horizontal and vertical momentum. 

\subsection*{Testing several collision conditions}
A vast experimental campaign was conducted, including three balls, two temperatures, two incident angles, for a range of impact speeds and initial contact locations. Furthermore, for one test we added a small amount of oil to the wall to lubricate the contact to induce sliding on the wall upon contact. The nick conditions were identified to occur for $0.6<H^*<0.75$ and $\tau<1$. All experiments conducted in this investigation satisfied these conditions, as shown, as shown in Fig. 2 of the main document. 

\subsubsection*{High speed collisions} 
An interesting group of data is that for $\tau<1$ but $H^*>0.75$ (data on lower right hand side of Fig. 2(b)). One would expect the condition of $\tau<1$ to be sufficient since it quantifies the ratio of rolling to contact time. 

It is not possible for a ball to reach the horizontal wall for $H^*>0.78$, if one considers that the maximum possible stretching on the wall upon impact (see above). However, the calculation from Eqn. 2, from the main text, results in values with $\tau<1$. We note that these experiments have small values of the Cauchy number (Ca $<0.16$) which correspond for high impact speeds ($U_0\approx 45$ m/s). For such a range of speeds, the deformation of the ball is significant. We do not expect of the calculation of the contact and rolling times to be valid in such a case. 

\subsubsection*{Experiments at different temperatures} 
For the experiments at $\theta_0=30$\textdegree, tests were conducted with the same three balls but at a higher temperature (40\textdegree C), to replicate  the conditions observed in a real squash game. At higher temperatures the elastic properties of the ball change (see Table 1 of the main document), leading to an increase in the coefficient of restitution \cite{berencsi2021}. The material becomes softer at higher temperatures (the value of $E$ decreases). The results for these tests are shown in Fig. 2(b) of the main text by the red symbols. These data also satisfy the nick occurrence conditions. Since, in general, $E$ decreases with temperature, the attainable values of Ca are smaller. 

\subsubsection*{Collisions over lubricated walls} 
An additional test was conduced  for the experiments at $\theta_0=25$\textdegree \ to further test our rolling hypothesis. For a set of experiments involving the double-dot and doubles balls, at room temperature, the vertical wall was covered by a thin film of viscous fluid (glycerol) with the intention to lower the surface friction. It is well known that when the wall is lubricated by a viscous fluid, sliding (instead of rolling) occurs upon contact\cite{joseph2004}. The results from these tests are  are shown by the green symbols in Fig. 2(a) of the main text.  The nick shot was prevented from occurring when the collision corresponded to a value of $\tau<1$. Upon closer inspection of the high speed videos, for these collisions the ball slid upon contact; hence, the rolling frustration mechanisms did not occur. 

\section*{Collision videos}
Videos for typical ball collisions have been placed on a data repository. All videos were recorded at 5000 fps, with a shuttle speed of 1/20000, using a Photron FASTCAM SA5 camera.\\

%%% Add this line AFTER all your figures and tables
%\FloatBarrier

\noindent \textbf{Movie S1:} {Case C1: Nick collision. \href{https://doi.org/10.26300/dgqx-9q58}{HERE}\\
    Conditions: Ca = 0.25, $H^*$ = 0.62, $\theta_0$ = 24.43$^o$ and $\tau$ = 0.87. }

\noindent \textbf{Movie S2: }{Case C2: No nick collision. \href{https://doi.org/10.26300/110c-cv13}{HERE}\\
    Conditions: Ca = 0.24, $H^*$ = 0.51, $\theta_0$ = 24.34\textdegree, and $\tau$ = 0.76.}
    
\noindent \textbf{Movie S3:} {Case C3: No nick collision. \href{https://doi.org/10.26300/xcmt-qm88}{HERE}\\
    Conditions: Ca = 0.25, $H^*$ = 0.81, $\theta_0$= 24.31$^o$ , and $\tau$ = 1.10.}

% Comment this out for submission
% \input{Supplemental}

\end{document}